\documentclass[twocolumn,letterpaper,prl,superscriptaddress,showpacs]{revtex4-1}
\usepackage{color}
\usepackage{amsmath}
\usepackage{graphicx}
\usepackage{epsfig}
\usepackage{subfigure}
\usepackage{mathrsfs}

\begin{document}

\title{Manipulating $Z_{2}$ and Chern topological phases in a single
material using periodically driving fields}
\author{Shu-Ting Pi}
\email{spi@ucdavis.edu}
\author{Sergey Savrasov}
\email{savrasov@physics.ucdavis.edu}
\affiliation{Department of Physics, University of California, Davis, One Shields Avenue, Davis, California 95616 USA}

\begin{abstract}
$Z_{2}$ and Chern topological phases such as newly discovered quantum spin
Hall and original quantum Hall states hardly both co--exist in a single
material due to their contradictory requirement on the time--reversal
symmetry (TRS). We show that although the TRS is broken in systems with a
periodically driving ac-field, an effective TRS can still be defined
provided the ac--field is linearly polarized or certain other conditions are
satisfied. The controllable TRS provides us with a route to manipulate $Z_{2}
$ and Chern topological phases in a single material by tuning the
polarization of the ac--field. To demonstrate the idea, we consider a
generic honeycomb lattice model as a benchmark system that is relevant to
electronic structures of several monolayered materials. Our calculation
shows that not only the transitions between $Z_{2}$ and Chern phases can be
induced but also features such as the dispersion of the edge states can be
controlled. This opens the possibility of manipulating various topological
phases in a single material and can be a promising approach to engineer some
new electronic states of matter.
\end{abstract}

\maketitle

\textit{Introduction.} The discovery of topological insulators (TIs) in
condensed matter systems has not only revealed novel physics of the quantum
world but also unified many physical phenomena, which were thought to be
irreverent, into the same framework\cite{TI-6}. Their peculiar edge states
make TIs a hot topic for both fundamental interests and industrial
applications. Several materials such as HgTe/CdTe quantum well, Bi$_{x}$Sb$%
_{1-x}$ alloys, Bi$_{2}$Se$_{3}$ and Bi$_{2}$Te$_{3}$, etc., have been
proven to be TIs by experiments\cite{TI-1}. Despite these successes, how to
design a topologically non--trivial material, remains a challenging issue.
In most cases, the discovery of new TIs still relies on serendipity rather
than predetermination.

Instead of searching for materials with intrinsically non--trivial topology,
there are several recent studies focusing on manipulation of topological
phases using controllable physical processes, e.g. electric fields, strains,
etc\cite{TT-1,TT-2,TT-3}. Those studies not only offer new tools to generate
various topological phases but also open new ways to making real electronic
devices.

One of the promising methods to engineer a topological property of a system
is to use periodically driving fields\cite{FB-1,FB-2,FB-3,FB-4,FB-5}. The
proposal is based on the Floquet theory which states that the Hamiltonian of
a system with a time--dependent periodic potential can be mapped into an
effective static Hamiltonian, called the Floquet Hamiltonian. If the
(quasi)band structure of a Floquet Hamiltonian exhibits a topological
behavior, we can expect there exists a similar feature in the original
Hamiltonian in a dynamical fashion. An advantage of using this method to
engineer the band topology is that the ac--field provides a set of tunable
parameters such that a variety of band structures unaccessible in the
original material can be generated in a dynamical way. Many proposals based
on the topology of Floquet Hamiltonians have appeared recently, some of
which are: Floquet TIs in graphene\cite{FTI-1}, Floquet TIs in semiconductor
quantum wells\cite{FTI-2}, Floquet Majorana fermions in topological
superconductors\cite{FTI-3}, merging Floquet Dirac points\cite{FTI-4},
Floquet fractional Chern insulators\cite{FTI-5}, Floquet Wely semimetal\cite%
{FTI-6}, etc. A few experiments that support the idea of Floquet TIs have
also\ been carried out\cite{FEX-1,FEX-2}. Those works not only lighten up
the road to manipulate topological phases but also bring us a vast landscape
of new physical phenomena that are hardly found in static systems.

While many topological phases have been studied within the Floquet
framework, the discussion of $Z_{2}$ phases remains scarce because
time--reversal symmetry (TRS), a necessary condition for the existence of
the $Z_{2}$ phase, is always broken due to the time dependence of the
external perturbation. However, the Floquet Hamiltonian is merely an
effective mapping of the original Hamiltonian, so the loss of TRS in
original Hamiltonian does not necessarily result in the loss of TRS in
Floquet Hamiltonian. Establishing an operator that links Floquet states in
the Brillouin zone by a similar way as conventional TR operator does it for
Bloch states, an effective TRS can be defined\cite{FTI-2,FB-2}. If so, two
seemingly contradictory phases, TRS protected $Z_{2}$ phase, such as
recently discovered quantum spin Hall state, and TRS broken Chern phase,
such as much celebrated original quantum Hall state, can both be manipulated
in a single material by tuning the ac--field, which is the main message of
the present work.

Here, we first show how we truncate the Floquet Hamiltoian to finite
dimension in a realistic calculation. Second, we show that the TRS
conditions can be easily satisfied if the field is linearly polarized or
certain low excitation conditions are reached. Third, we use a prototypical
2D material with strong spin--orbit coupling as a benchmark in our
calculation, in order to demonstrate the idea of manipulating $Z_{2}$ and
Chern topological phases in the same system. More specifically, we consider
a generic half--filled $p$--orbital honeycomb lattice model to illustrate
our findings. Our results show the evidence for the $Z_{2}$ phase with the
ability to control the dispersion of its edge states by properly tailoring
the external ac--field. When the polarization is away from the effective TRS
condition, a rich Chern phase diagram begins to appear which suggests a Z$%
_{2}$--Chern phase transition. Thus we demonstrate the possibility of
manipulating between the two topological phases in a single material which
can serve as a promising tool to engineer some novel electronic states in
condensed matter systems.

\textit{Floquet Theorem.} We consider a tight--binding Hamiltonian with an
external time perodical ac--field $H(\tau )=\sum_{\alpha \beta
}\sum_{jl}t_{jl}^{\alpha \beta }(\tau )c_{\beta l}^{\dagger }(\tau
)c_{\alpha j}(\tau )+h.c.$ where $\tau $ is time, $(\alpha ,\beta )$ are the
internal degrees of freedom (e.g. orbitals, spins, etc.) of the unit cell
positioned at $\mathbf{R}_{j}$ and $\mathbf{R}_{l}$. The ac--field is
coupled to the problem by introducing a minimal coupling $t_{jl}^{\alpha
\beta }(\tau )\rightarrow t_{jl}^{\alpha \beta }e^{i\mathbf{A}(\tau )(%
\mathbf{r}_{j}^{\alpha }-\mathbf{r}_{l}^{\beta })}$ where $\mathbf{r}%
_{j}^{\alpha }$ is the position vector of the state $|\alpha \rangle $ in
the unit cell located at $\mathbf{R}_{j}$ and $\mathbf{A}$ is the vector
potential of the field. Since the Hamiltonian has both lattice and time
translational symmetries, we can perform a dual Foruier transform , such as $%
c_{\alpha j}(\tau )=\sum_{n}c_{\alpha jn}e^{-in\omega \tau }=N^{-D/2}\sum_{%
\mathbf{k}}\sum_{n}c_{\alpha n}(\mathbf{k})e^{-i(\mathbf{k}\cdot \mathbf{R}%
_{j}+n\omega \tau )}$. The Floquet theorem proves that the Floquet
Hamiltonian $H_{F}(\tau )=H(\tau )-i\partial _{\tau }$ in the Fourier
transformed $(\mathbf{k},\omega )$ space can be expressed as\cite{FTI-5} 
\begin{align}
H_{F}(\mathbf{k},\omega )& =\sum_{nm\alpha \beta }(h_{nm}^{\alpha \beta
}-n\omega \delta _{nm}\delta _{\alpha \beta })c_{\alpha n}^{\dagger }(%
\mathbf{k})c_{\beta m}(\mathbf{k})+h.c.,  \notag  \label{Hf} \\
h_{nm}^{\alpha \beta }(\mathbf{k})& =\sum_{l}[t_{0l}^{\alpha \beta }J_{m-n}(%
\mathbf{A}(\tau )\cdot \Delta \mathbf{r})]e^{i\mathbf{k}\cdot \mathbf{R}%
_{l}}, \\
J_{q}(x(\tau ))& =\frac{1}{T}\int_{0}^{T}e^{i(x(\tau )-q\omega \tau )}d\tau ,
\notag
\end{align}%
where $\Delta \mathbf{r}=\mathbf{r}_{0}^{\alpha }-\mathbf{r}_{l}^{\beta }$, $%
\mathbf{k}$ is the wave vector, $\omega =2\pi /T$ is the frequency of the
ac--field and $(n,m)$ are the Floquet indexes.

Because Eq.\ref{Hf} is a key result of the Floquet theorem, we assert here
that 1) Similar to the undriven system, the Floquet Hamiltonian $H_{F}$
forms an eigenvalue problem $H_{F}(\mathbf{k},\omega )|u_{\gamma n}(\mathbf{k%
})\rangle =\epsilon _{\gamma n}(\mathbf{k})|u_{\gamma n}(\mathbf{k})\rangle $
where $\gamma $ is the band index, $n$ is the Floquet index ranging $-\infty 
$ to $+\infty $ and $\epsilon _{\gamma n}$ is the so called quasienergy; 2)
Similar to the existence of reciprocal lattice and the periodicity in the
k--space, the relations $\epsilon _{\gamma n}=\epsilon _{\gamma 0}+n\omega $
and $|u_{\gamma n}\rangle =|u_{\gamma 0}\rangle $ are held as a result of
the analogous properties of the Brillouin zone in the frequency domain. They
also show the physics of absorbing/emitting $n$ photons, so the Floquet
bands are shifted by $\pm n\omega $; 3) The solution of the original
Hamiltonian is obtained by linearly combining static Floquet band states $%
|\psi _{\gamma }(\tau )\rangle =e^{-i\epsilon _{\gamma }\tau }|u_{\gamma
}(\tau )\rangle =e^{-i\epsilon _{\gamma }\tau }\sum_{n=-\infty }^{+\infty
}e^{in\omega \tau }|u_{\gamma n}\rangle $ where $|u_{\gamma }(\tau )\rangle $
is the Floquet state which is periodic both in space and time. Note that $%
\tau $ no longer appears in $H_{F}$ and $|u_{\gamma n}\rangle $, so the
Floquet theorem simplifies the original time--dependent problem by mapping
it to a static one. Therefore we can treat $H_{F}$ as the usual lattice
Hamiltonian and explore its topology using the techniques developed for
static systems. If $H_{F}$ has non--trival edge states, we can expect a
dynamical analogy on $|\psi _{\gamma }(\tau )\rangle $\cite{FTI-2}; 4) The
form of $h_{nm}^{\alpha \beta }(\mathbf{k})$ is just the usual lattice
Fourier transform of the states labeled by two indexes $(\alpha ,n)$ rather
than by one. The extra degree of freedom is the penalty of mapping the
time--dependent Hamiltonian into a static one. The hopping integrals between
the two states $c_{\alpha jn}^{\dagger }$ and $c_{\beta lm}$ are obtained by
modifying the undriven hoppings: $t_{jl}^{\alpha \beta }\rightarrow
t_{jl}^{\alpha \beta }J_{m-n}(\mathbf{A}\cdot \Delta \mathbf{r})$. 

Because the Floquet index $n$ ranges from $-\infty $ to $+\infty $, the
Floquet Hamiltonian is not manageable unless we make some approximations\cite%
{FB-1}. Two approximations are frequently adopted: (a) weak intensity limit
and (b) high--frequency limit.

\begin{figure}[tbp]
\centering
\includegraphics[width=1.0\columnwidth]{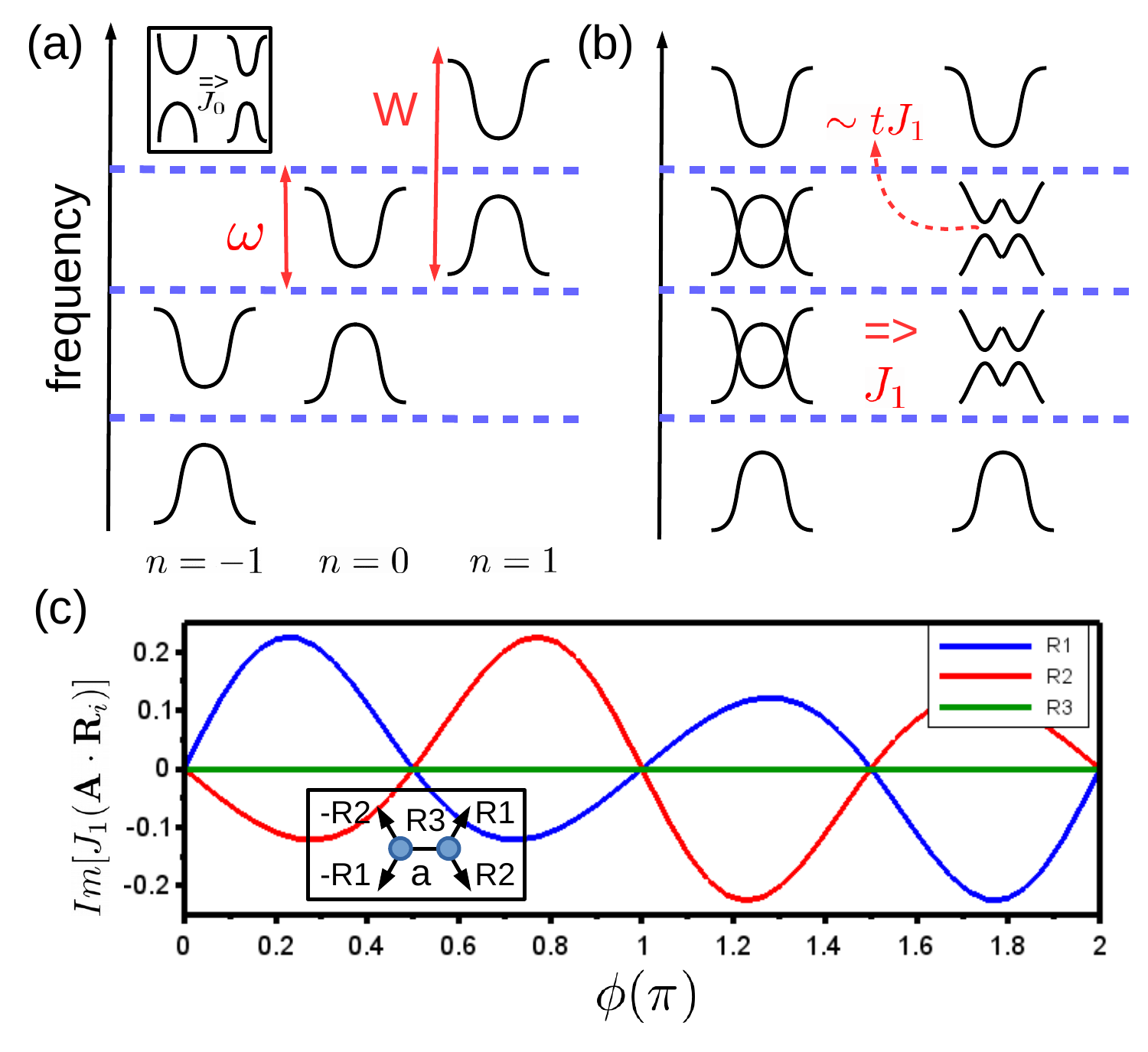}
\caption{(color) (a) Floquet bands within first order emission/absorption
photon processes as replicas of the original band structure modified by $%
J_{0}$ (see left upper inset). (b) formation of the Floquet band structure
by merging states into single Brillouin zone (left) and accounting for the
effect of gap opening due to $J_{1}$ (right). (c) The imaginary part of $%
J_{1}(\mathbf{A}\cdot \mathbf{R}_{j})$ as a function of polarization $%
\protect\phi $ (in unit of $\protect\pi $) when $\mathbf{A}=[1,7.467]/a$
(with $\hbar =e=1$). Left lower inset gives the definition of each $\mathbf{R%
}_{i}$ of graphene honeycomb lattice, where $a$ is the length of $\mathbf{R}%
_{3}$. }
\end{figure}

For the approximation (a), let us consider an ac--field sinusoidal in time.
In this case, $J_{q}(\mathbf{A}\cdot \Delta \mathbf{r})$ is essentially the $%
q$--th Bessel function of the first kind. In the limit of the weak
intensity, $|A|\rightarrow 0$, its asymptotic behavior is as follows: $%
J_{0}\rightarrow 1,$ $J_{q\neq 0}\rightarrow 0$. The larger the $q$ the
faster $J_{q\neq 0}$ drops to zero. Hence we can truncate $H_{F}$ to a
finite dimension by including just a few lowest order photon processes,
provided the field intensity is weak enough. For example, if we keep $q=0,1$%
, $H_{F}$ is reduced to the following form 
\begin{align*}
H_{F}& \simeq \mathcal{P}_{1}H_{F}\mathcal{P}_{1}^{-1}=H_{f}^{1}= \\
& \left( 
\begin{array}{ccc}
h_{0}^{\alpha \beta }-1\omega I & h_{1}^{\alpha \beta } & 0 \\ 
h_{-1}^{\alpha \beta } & h_{0}^{\alpha \beta } & h_{1}^{\alpha \beta } \\ 
0 & h_{-1}^{\alpha \beta } & h_{0}^{\alpha \beta }+1\omega I%
\end{array}%
\right) ,
\end{align*}%
where $H_{f}^{1}$ denotes a reduced Floquet Hamiltonian describing an
emission/absorbtion of a single photon and $\mathcal{P}_{1}$ is the operator
that projects $H_{F}$ to $H_{f}^{1}$ . A diagrammatic explanation of such
first order process is shown in Fig.1(a) and (b). In the upper left inset,
the undriven band structure is modified by the 0--th order effect $J_{0}$.
Once $J_{1}$ term comes in, the bands will have three copies with energy
shifts $0,\pm \omega $. When those bands reach resonant energies, i.e the
band crossings, $J_{1}$ will open gaps $\sim tJ_{1}$ making them
anti--crossing. This is the main idea of the truncation. One has to note
that when $H_{F}$ is truncated, the periodicity in frequency domain is
broken,\ and the relations $\epsilon _{\gamma n}=\epsilon _{\gamma
0}+n\omega $ and $|u_{\gamma n}\rangle =|u_{\gamma 0}\rangle $ are strictly
speaking no longer valid.

As for the approximation (b), let us assume the frequency of the external
field is so much larger than the bandwidth, $\omega \gg W,$ that the Floquet
bands do not cross anymore. In this limit, the gap openings due to $J_{q>0}$
become less important, which implies that it is also the condition to
consider just the lowest order photon processes.

\textit{Time-Reversal Symmetry.} In an undriven system, the TRS is defined
by $\mathcal{T}H(\tau )\mathcal{T}^{-1}=H(-\tau )$ where $\mathcal{T}$ is
the conventional TR operator $\mathcal{T}=e^{-i\pi \sigma _{y}/2}K$.
Although systems with time--dependent ac--fields do not hold this property,
it is still possible to define an effective TRS for the Floquet Hamiltonian%
\cite{FTI-2,FB-2}. To give specific conditions holding the effective TRS, we
conclude with two important theorems here (see Supplementary Materials):

\textbf{Theorem\ I:} If there exists a parameter $\tau _{0}$ such that $%
\mathcal{T}H(\tau )\mathcal{T}^{-1}=H(\tau +\tau _{0})$, one can always
define an effective TR operator $Q=e^{iH_{F}\tau _{0}}\mathcal{T}$ that
satisfies the relation $\mathcal{Q}H_{F}(\mathbf{k})\mathcal{Q}^{-1}=H_{F}(-%
\mathbf{k}).$

\textbf{Theorem\ II:} Assuming a system has TRS when it is undriven, i.e. $|%
\mathbf{A}|=0$, then $\mathbf{A}(\tau)=[A_{x}sin(\omega
\tau+\phi_{x}),A_{y}sin(\omega \tau+\phi_{y}),A_{z}sin(\omega \tau+\phi_{z})]
$ with $\phi_{i}-\phi_{j}=m\pi$ ($i,j \in\ x,y,z$; $m \in integers$) will
automatically make $H(\tau)$ satisfy $\mathcal{T}H(\tau)\mathcal{T}%
^{-1}=H(\tau+\tau_{0})$. Furthermore if the time frame is properly chosen,
one can always let all $\phi_{i}^{\prime }s=n_{i}\pi\ (n_{i} \in integer)$
such that $\tau_{0}=0$ and $Q=\mathcal{I}\mathcal{T}$

These theorems tell us if the phase differences among each field component
are multiples of $\pi $, the Floquet Hamiltonian will have effective TRS\cite%
{FTI-6} and the TR operator can be treated as a conventional one acting in
the Hilbert space of the basis of the Floquet Hamiltonian $\{|\alpha n(%
\mathbf{k})\rangle \}$. In the following, we will call the condition $\phi
_{i}-\phi _{j}=m\pi $ as a linear polarization although the polarization is
not definable if $\mathbf{A}$ is not in 2D.

The linear polarization condition is not the only option to have effective
TRS. Since we are handling the $\nu $--th order reduced Floquet Hamiltonian $%
H_{f}^{\nu }$ rather than the original $H_{F}$ in a realistic calculation,
it is possible that $H_{f}^{\nu }$ has more time--reversal points than $H_{F}
$. Recall that the hopping integrals in the Floquet Hamiltonian are
generated by modifying $t_{\alpha \beta }^{ij}\rightarrow t_{\alpha \beta
}^{ij}J_{q}(\mathbf{A}(\tau )\cdot \Delta \mathbf{r})$. If one can properly
tailor $\mathbf{A}$ such that $J_{q\leq \nu }$ are real numbers for all $%
\Delta \mathbf{r}$ in the lattice, obviously TRS will be kept up to $\nu $%
--th order $H_{f}^{\nu }$. Since $J_{0}$ is always real, the highest order $%
\nu $ should be equal or greater than 0. To give an example of $\nu >0$, we
have plotted in Fig. 1(c) the imaginary part of $J_{1}$ with respect to
three non--equivalent position vectors of a honeycomb lattice as a function
of polarization $\phi =\phi _{x}-\phi _{y}$ by setting $%
[A_{x},A_{y}]=[1,7.467]/a$. One can immediately notice that there are two
additional TRS points (all lines reach 0) other than $\phi =m\pi $, i.e. $%
\phi =\pi /2$ and $\phi =3\pi /2$. However, those TRS points are just
results of low excitation approximation. One should always confirm that the
energy splitting $\Delta E$ of Kramer degenerate states due to higher order
terms is much smaller than the characteristic energy $\varepsilon _{c}$ that
we are interested in ($\Delta E\sim tJ_{\nu +1}\ll \varepsilon _{c}$) to
explore this feature further.

\begin{figure}[tbp]
\centering
\includegraphics[width=1.0\columnwidth]{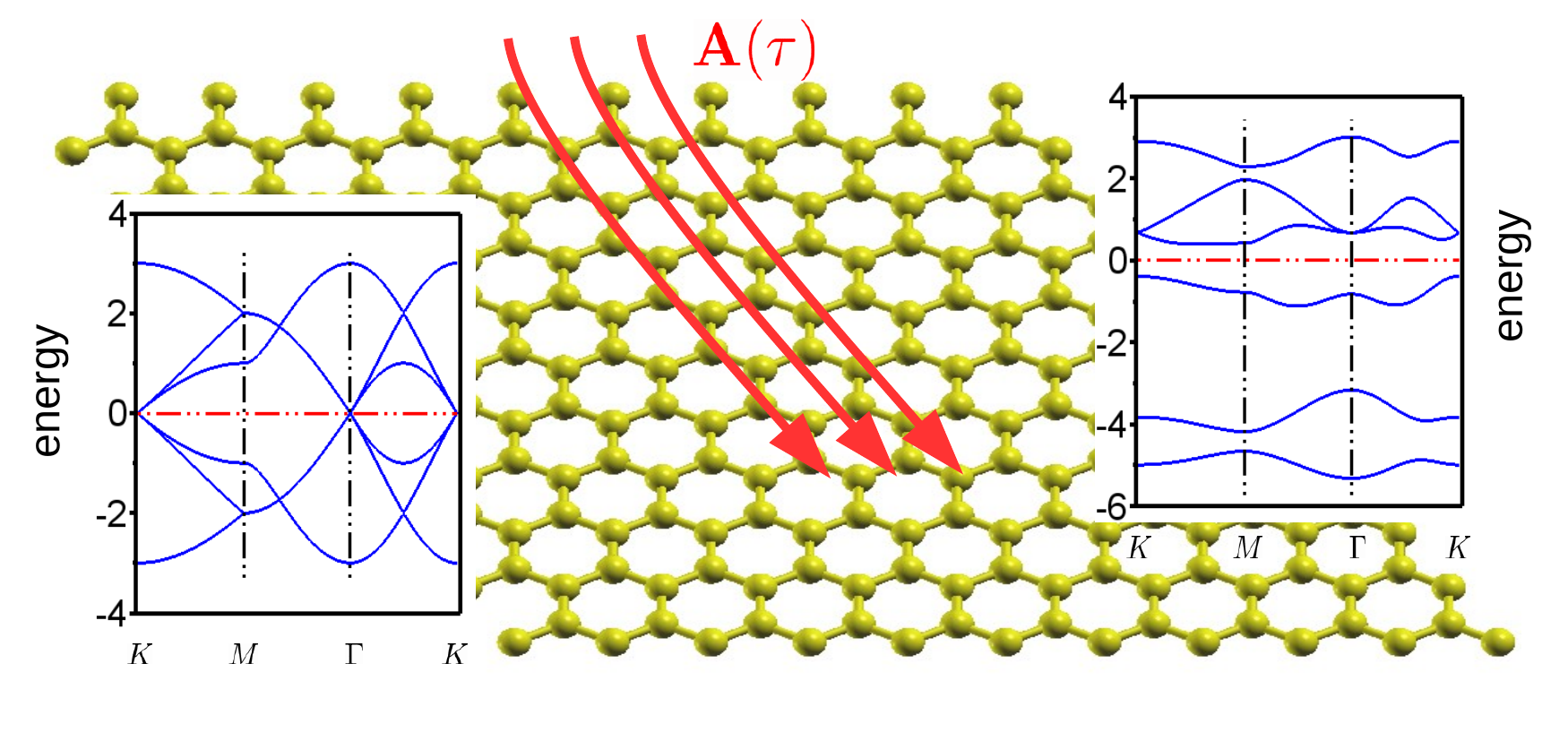}
\caption{(color) Honeycomb lattice irradiated by electric ac-field $\mathbf{A%
}(\protect\tau )$. Left inset: band structure without SOC. Right inset: band
structure with SOC $\protect\lambda =3t$, $\ $\ $t$ being a unit of energy.
Note that all the bands are doubly degenerate.}
\end{figure}

\textit{Floquet Topological Phases.} The best candidates to realize the
transition between a TRS Floquet $Z_{2}$ phase and a non--TRS Floquet Chern
phase would be 2D materials with spin--orbit coupling, e.g transit--metal
dichalcogenides\cite{TMD-1}, graphene with adatoms\cite{GAD-1,GAD-2},
silicene\cite{SIL-1}, germanane\cite{GER-1}, Tin films\cite{TIN-1}, $\alpha $%
-Sn\cite{ASN-1}, etc. These materials have been proven (or have high
expectancy) to exhibit monolayer structures with band gaps around dozens to
hundreds meV. Because of their planar geometry, the in--plane ac--field can
be easily realized by a laser in experimental setup.

Here we\ consider a nearest neighbor tight--binding Hamiltonian on a
honeycomb lattice with a $p$--orbital (total six states) per each site as a
generic minimal model describing the 2D material at the center of interest.
For simplicity, we assume the system is half--filled. In order to make our
model close to actual band structures, hopping integrals are generated by a
Slater--Koster method\cite{SKI-1} with $V_{pp\sigma }=t$, $V_{pp\pi }=-t$
and onsite energy $E_{p_{x,y,z}}=0$. SOC is treated as a local potential by
evaluating the matrix elements $\langle p_{i}|\lambda \mathbf{L}\cdot 
\mathbf{S}|p_{j}\rangle $ with $\lambda =3t$ for each site. In order to
calculate the topological invariants, we implement the $n$--field method
introduced Fukui et al.\cite{TII-1}. This method has been proven to provide
evaluations of both $Z_{2}$ and Chern topological invariants in discretized
Brillouin zones accurately and efficiently. We emphasize extra time that
when computing $Z_{2}$ invariants for the Floquet Hamiltonian, the TR
operator should be replaced by the effective TR operator $Q=e^{iH_{F}\tau
_{0}}\mathcal{T}$ \ as described in this work.

\begin{figure}[tbp]
\centering
\includegraphics[width=1.0\columnwidth]{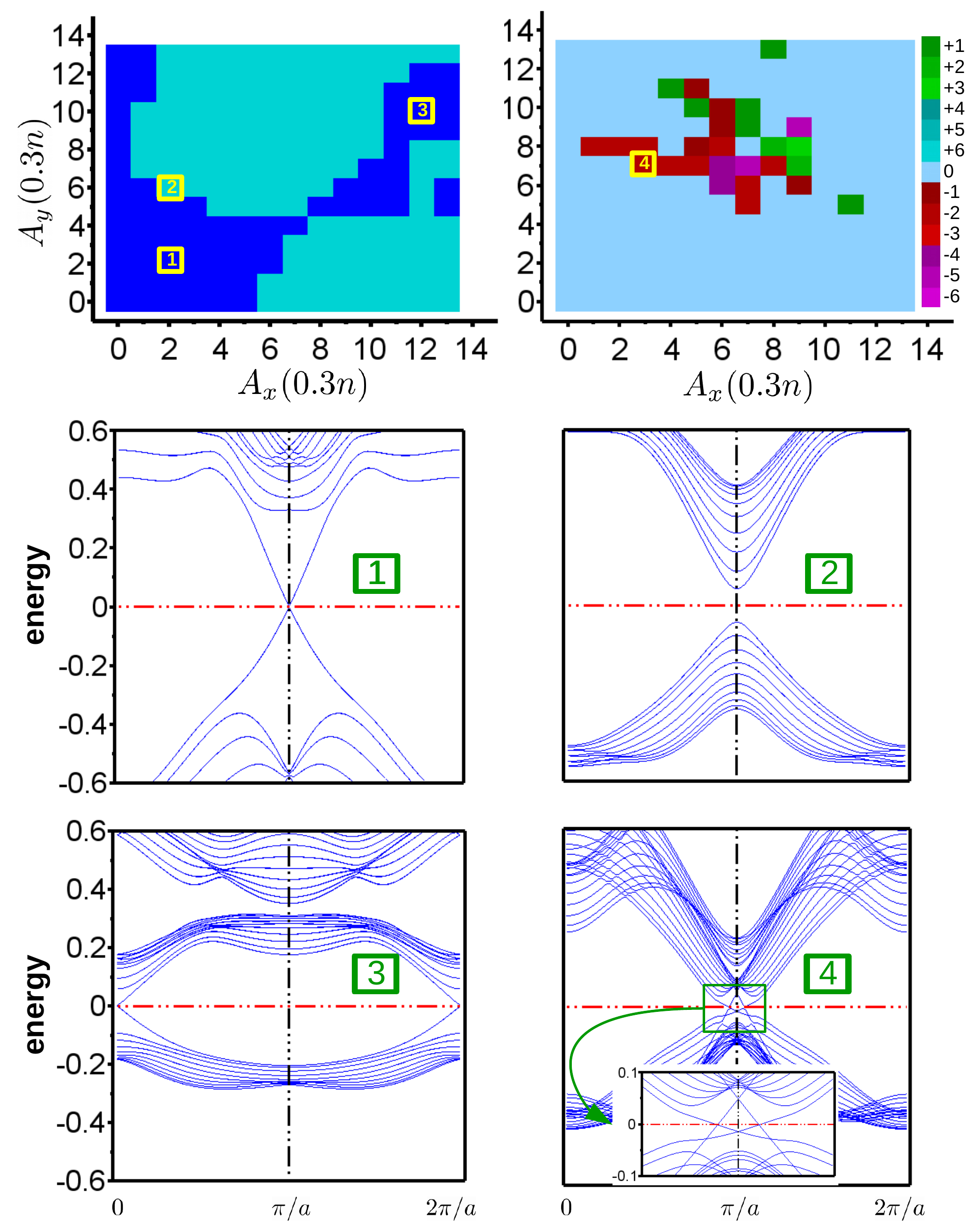}
\caption{(color) Floquet topological phase diagram. Upper left: $Z_{2}$
phase diagram (green for $Z_{2}=0$ and blue for $Z_{2}=1$) with linearly
polarized ac--field $[\protect\phi _{x},\protect\phi _{y}]=[0,0]$. Upper
right: Chern phase diagram (Chern numbers for each color are shown in the
legend) with eliptically polarized ac--field $[\protect\phi _{x},\protect%
\phi _{y}]=[0.0,0.5]\protect\pi $. $A_{x}$ and $A_{y}$ are chosen to be $%
0.3n/a,\ n=1\sim 13$ ($a$ is lattice constant, $\hbar =e=1$). There are four
points at the phase diagram labeled by $1\ $to $4$ with their edge--state
band structures shown (all energies in units of $t$) . Edge states are
calculated in a zigzag ribbon geometry with $20$ sites in the transverse
direction.}
\end{figure}

In Fig.2, we show a cartoon of the honeycomb lattice and the band structures
with and without SOC. One can find that the Dirac points at high symmetry
points $\Gamma $ and $K$ become gapped when SOC is turned on. This is a
general feature of a system with the honeycomb lattice. To study the Floquet
effect, we consider the reduced Floquet Hamiltonian to first order, $%
H_{f}^{1},$ and use a rather large frequency $\omega =10t$ (larger than the
band width $\sim 8t$). The amplitudes of $A_{x}$ and $A_{y}$ are chosen to
be $0.3n/a,\ n=1\sim 13$ ($a$ is the lattice constant) with linearly
polarized field $\phi _{x}=\phi _{y}=0$. The Floquet bands are also assumed
to be half--filled as in the undriven case. Fig.3 shows the $Z_{2}$ phase
diagram and selected band structures (labeled by $1$ to $4$) of the edge
states. Apparently there exists a large (shown in blue) area of $Z_{2}$
phase in the parameter space. To check the corresponding edge states, we
have plotted the Floquet band structures of zigzag ribbons under the same
ac--field. Plots corresponding to the parameters of phase points 1 and 3 are
both $Z_{2}=1$ phases, so the Dirac cones appear in the gapped region as
expected. An interesting finding for the phase point 3 is that the Dirac
cone appears in $k=0$ rather than at $k=\pi /a$ as seen for the point 1. It
is supposed to be the case of the armchair ribbon in the Kane-Mele model
which now appears in the zigzag edge with appropriate parameters\cite{KMM-1}%
. It means the ac--field can not only tune the $Z_{2}$ topology from trivial
to non--trivial, but also control the specific features of the edge states.
Finally, we have also plotted the band structure corresponding to the
trivial phase (point 2) as a confirmation that the edge Dirac cone indeed
does not show up.

Now we discuss how to make the $Z_{2}$ phase transiting to a Chern phase.
Let us consider an elliptically (circularly if $A_{x}=A_{{y}}$) polarized
ac--field with $[\phi _{x},\phi _{y}]=[0,\pi /2]$. The phase diagram of the
Chern numbers is shown in the upper right of Fig.3. Because it is a
multiband problem (12 bands for the undriven and 36 bands for the reduced
Floquet Hamiltonian), the Chern number can be much larger than $\pm 1$\cite%
{MTB-1}. We also show the edge states corresponding to the parameters of the
phase point 4. The existence of two Dirac cones at the edge agrees with our $%
C=-2$ result very well. The phase diagrams of Fig.3 illustrate how $Z_{2}$
and Chern topological phases can be manipulated in a single material using
properly tailored ac--fields.

Finally, we estimate some physical quantities relevant to realization of
such exotic electronic phases in real systems. Let us take graphene with
adatoms as an example\cite{GAD-1}. It is predicted to have SOC induced gap $%
E_{g}$ around $5\sim 20$ meV. To simulate this problem, we use
tight--binding parameters for the $s\ $and $p$ states of graphene obtained
by fitting to its band structure\cite{GTB-1} and tune the SOC to a value
that in our model fits the gap value of 5 meV. We consider two cases: a
microwave field, $\omega =2GHz\ll E_{g}/\hbar ,$ and an infrared field, $%
\omega =2THz\simeq E_{g}/\hbar $. Polarization angles are chosen to be $0$
and $\pi $ in order to observe $Z_{2}$ and Chern phases respectively. To
ensure the weak intensity approximation, we limit $A_{x},A_{y}<1(\hbar /%
\mathring{A})$ for both cases so that $J_{2}$ effects are about two orders
of magnitude smaller that $J_{1}$ and can be neglected. The electric field
and the corresponding laser intensity are obtained by $E_{0}=A\omega /e$ and 
$I_{0}=\frac{1}{2}\epsilon _{0}cE_{0}^{2}=1.33\times
10^{-3}E_{0}^{2}(W/cm^{2})$ respectively. For the microwave case, we found
that the Floquet $Z_{2}$ phase can be easily observed but the Chern phase
cannot. This corresponds to the electric fields $E_{0}<131V/cm$ or the
intensities $I_{0}<23W/cm^{2}$, which can be achieved by lasers with powers $%
P\lesssim 200mW$, easily accessible in experiment. As for the infrared case,
we found both $Z_{2}$ and Chern phases can be generated within that regime.
It corresponds to the electric fields $E_{0}<1.31\times 10^{5}V/cm$ or the
intensities $I_{0}<2.3\times 10^{7}W/cm^{2}$. This will require a rather
high power about several kW in experiments. This power is still
experimentally accessible but most materials can burn out under such a
strong field. Therefore searching for a material that can display both
phases under lower intensities could be an interesting topic for future
research. Although realization of $Z_{2}$--to--Chern phase transition could
be difficult in experiments for our discussed system, we have to emphasize
that easily achievable Floquet $Z_{2}$ phase still remains a treasure in
problems of engineering topological electronic structures.

\textit{Conclusion.} In summary, we have developed a framework to study TRS
in Floquet Hamiltonian and used a generic tight--binding model of the
honeycomb lattice relevant to several recently discovered monolyaered
materials in order to demonstrate the transition between $Z_{2}$ and Chern
phases by tuning the polarization of the ac--field. Although, our discussion
is based on the dynamical analogies, the physics is still very fascinating
not only due to the emergence of the $Z_{2}$ phase in a formally
time--reversal breaking potential but also due to the possibility of
manipulating self--contradictory topological phases in a single material.
Both phenomena are hard to find in static systems but could lead us to a new
physics that is unreachable in conventional solid--state matter.

\textit{Acknowledgments} We would like to acknowledge the useful discussions
with X. Dai, X. Wan, R. Wang and B. Wang. We also acknowledge the support by
NSF DMR Grant No.1411336.

\newpage
\section{Floquet Time-Reversal Symmetry}
Define Floquet operator $u(T)$ and Floquet Hamiltonian $H_{F}$ 
\begin{equation*}
u(T)=\mathbf{T}[e^{-i\int_{0}^{T}H(\tau )d\tau }]\equiv e^{-iH_{F}T}
\end{equation*}%
where $T$ is the time periodicity of the Floquet system and $\mathbf{T}$ is
the time-order product. We hope to find an effective time--reversal (TR)
operator $\mathcal{Q}$ for the Flouqet operator and Floquet Hamiltonian such
that 
\begin{equation*}
\mathcal{Q}u(T)\mathcal{Q}^{-1}=u(-T)
\end{equation*}%
and 
\begin{equation*}
\mathcal{Q}H_{F}(\mathbf{k})\mathcal{Q}^{-1}=H_{F}(-\mathbf{k})
\end{equation*}%
where $\mathcal{Q}$ is an antilinear oparator with $\mathcal{Q}^{2}=-1$. We
claim that if there exists an parameter $\tau _{0}$ that satisfies the
relation 
\begin{equation*}
\mathcal{T}H(\tau )\mathcal{T}^{-1}=H(-\tau +\tau _{0})
\end{equation*}%
where $\mathcal{T}$ is the conventional TR operator. Then an effective $%
\mathcal{Q}$ can always be defined as 
\begin{equation*}
\mathcal{Q}\equiv u(0,\tau _{0})\mathcal{T}=e^{iH_{F}\tau _{0}}\mathcal{T}
\end{equation*}%
In the following, we provide a proof for this theorem. (Note: Our proof is 
equivalent to the one shown in Ref.7 of the main article. Because 
we have chosen a slightly different statement, we prove it again here.)

Let us represent the conventional TR operator as the product of an unitary
operator $\mathcal{S}$ (usually $e^{-i\pi\sigma_{y}/2}$) and the complex
conjugate operator $\mathcal{K}$: 
\begin{equation*}
\mathcal{T}=\mathcal{S}\mathcal{K}
\end{equation*}
Assume there exists a parameter $\tau_{0}$ such that 
\begin{equation*}
\mathcal{T}H(\tau)\mathcal{T}^{-1}=H(-\tau+\tau_{0}) \Longleftrightarrow 
\mathcal{S}H^{\ast}(\tau)\mathcal{S}^{\dagger}=H^{\dagger}(-\tau+\tau_{0})
\end{equation*}
Since 
\begin{align*}
u(T,0)&=\lim_{N \rightarrow \infty} e^{-i \Delta \tau H(T-\Delta \tau)}
\times e^{-i \Delta \tau H(T-2\Delta \tau)} \times \cdots \\
& \cdots \times e^{-i \Delta \tau H(0)}\ \ ; \ \ \Delta \tau=T/N
\end{align*}
We have 
\begin{align*}
\mathcal{S}u^{*}(T,0)\mathcal{S}^{\dagger}&=\lim_{N \rightarrow \infty}
e^{i\Delta \tau \mathcal{S} H^{*}(T-\Delta \tau) \mathcal{S}^{\dagger}}
\times \cdots \times e^{i\Delta \tau \mathcal{S} H^{*}(0) \mathcal{S}%
^{\dagger}} \\
&=\lim_{N \rightarrow \infty} e^{i\Delta \tau H^{\dagger}(\tau_{0}-T+\Delta
\tau)} \times \cdots \times e^{i\Delta \tau H^{\dagger}(\tau_{0})} \\
&=\lim_{N \rightarrow \infty} e^{i\Delta \tau H^{\dagger}(\tau_{0}+\Delta
\tau)} \times \cdots \times e^{i\Delta \tau H^{\dagger}(\tau_{0}+T)} \\
&=u^{\dagger}(\tau_{0}+T,\tau_{0})
\end{align*}
where the third equal sign uses the relation $H(\tau+T)=H(\tau)$. Therefore
if we define $\mathcal{R}\equiv u(0,\tau_{0})\mathcal{S}$ to shift the
origin from $\tau_{0}$ to 0, then an effective TR operator can be defined as 
$\mathcal{Q}=\mathcal{R}\mathcal{K}$ 
\begin{align*}
\mathcal{R}u^{*}(T,0)\mathcal{R}^{\dagger}&=u(0,\tau_{0}) \mathcal{S}%
u^{*}(\tau,0)\mathcal{S}^{\dagger} u^{\dagger}(0,\tau_{0}) \\
&=u(0,\tau_{0})u^{\dagger}(\tau_{0}+T,\tau_{0})u^{\dagger}(0,\tau_{0}) \\
&=u^{\dagger}(T,0)
\end{align*}
It means 
\begin{equation*}
\mathcal{Q}u(T,0)\mathcal{Q}^{-1}=u(-T,0)
\end{equation*}
and 
\begin{equation*}
\mathcal{Q}H_{F}(\mathbf{k})\mathcal{Q}^{-1}=H_{F}(-\mathbf{k})
\end{equation*}

\section{Relation to Polarization}

Consider a vector potential $\mathbf{A}(\tau )=[A_{x}sin(\omega \tau +\phi
_{x}),A_{y}sin(\omega \tau +\phi _{y}),A_{z}sin(\omega \tau +\phi _{z})]$.
We claim two consequences:

\begin{itemize}
\item If $\phi _{i}-\phi _{j}=m\pi $ ($i,j\in x,y,z$, $m\in integer$), the
Floquet time--reversal criterion: $\mathcal{T}H(\tau )\mathcal{T}%
^{-1}=H(-\tau +\tau _{0})$ will always be satisfied.

\item If $\phi_{i}-\phi_{j}=m\pi$, one can always let $\phi_{i}=n_{i}\pi$ ($%
n_{i}\in integer)$ such that $\tau_{0}=0$ and the effective TR operator can
be simply expressed as $\mathcal{Q}=\mathcal{IT}$.
\end{itemize}

The first theorem tells us the relation between the Floquet TR symmetry and
the polarization of the ac--field. The second theorem helps us to deal with
the effective TR operator in a much simpler way. In the following, we
provide a proof for these two statements.

Consider the basis set of Hilbert space $\{|\alpha ,\sigma \rangle \}$ where 
$\alpha $ is the label of space--related degree of freedom, e.g, sublattice,
orbital, etc., and $\sigma =+/-$ is the spin index. Define time--reversal
operator $\mathcal{T}=uK$ where $u=-i\sigma _{y}$. Then the matrix element
of a time--reversal transformation applied to the Hamiltonian is given by 
\begin{align*}
\langle \alpha \sigma |\mathcal{T}H\mathcal{T}^{-1}|\alpha ^{^{\prime
}}\sigma ^{^{\prime }}\rangle & =\langle \alpha \sigma |uH^{\ast }u^{\dagger
}|\alpha ^{^{\prime }}\sigma ^{^{\prime }}\rangle  \\
& =(-1)^{[\delta _{\sigma -}+\delta _{\sigma ^{^{\prime }}-}]}(H_{\alpha
^{^{\prime }}-\sigma ^{^{\prime }}}^{\alpha -\sigma })^{\ast }
\end{align*}%
If TR symmetry exists, $\mathcal{T}H\mathcal{T}^{-1}=H$, and we obtain a
restriction on the matrix elements: 
\begin{equation*}
(-1)^{[\delta _{\sigma -}+\delta _{\sigma ^{^{\prime }}-}]}(H_{\alpha
^{^{\prime }}-\sigma ^{^{\prime }}}^{\alpha -\sigma })^{\ast }=H_{\alpha
^{^{\prime }}\sigma ^{^{\prime }}}^{\alpha \sigma }
\end{equation*}%
For a system with an ac--field, the hopping integral is modified by $%
t_{\alpha ^{^{\prime }}\sigma ^{^{\prime }}}^{\alpha \sigma }(\tau
)\rightarrow t_{\alpha ^{^{\prime }}\sigma ^{^{\prime }}}^{\alpha \sigma
}e^{i\mathbf{A}(\tau )(\mathbf{r}_{\alpha }-\mathbf{r}_{\alpha ^{^{\prime
}}})}$. The Floquet TR criterion $\mathcal{T}H(\tau )\mathcal{T}%
^{-1}=H(-\tau +\tau _{0})$ requires that 
\begin{align*}
& (-1)^{[\delta _{\sigma -}+\delta _{\sigma ^{^{\prime }}-}]}(t_{\alpha
^{^{\prime }}-\sigma ^{^{\prime }}}^{\alpha -\sigma })^{\ast }e^{-i\mathbf{A}%
(\tau )(\mathbf{r}_{\alpha }-\mathbf{r}_{\alpha ^{^{\prime }}})} \\
& =t_{\alpha ^{^{\prime }}\sigma ^{^{\prime }}}^{\alpha \sigma }e^{i\mathbf{A%
}(-\tau +\tau _{0})(\mathbf{r}_{\alpha }-\mathbf{r}_{\alpha ^{^{\prime }}})}
\end{align*}

Assuming the system has TR symmetry when undriven, the hopping integrals
will be canceled out and we have 
\begin{equation}\label{A_crit}
-\mathbf{A}(\tau )=\mathbf{A}(-\tau +\tau _{0})
\end{equation}%
Because 
\begin{equation*}
\mathbf{A}(\tau )=[A_{x}sin(\omega \tau +\phi _{x}),A_{y}sin(\omega \tau
+\phi _{y}),A_{z}sin(\omega \tau +\phi _{z})]
\end{equation*}%
then Eq.\ref{A_crit} means 
\begin{align} \label{sine_eq}
& -(\omega \tau +\phi _{i})+2n_{i}\pi =-\omega \tau +\omega \tau _{0}+\phi
_{x}   \\
\Rightarrow & \ \phi _{i}=n_{i}\pi -\omega \tau _{0}/2,\ \ \ i\in x,y,z\ ;\
n_{i}\in integer  \notag
\end{align}%
Since $\tau _{0}$ can be arbitrary real numbers, it is convenient to express
the effective TR condition as 
\begin{equation} \label{linear}
\phi _{i}-\phi _{j}=m\pi ,\ \ \ m\in integer
\end{equation}%
For an in--plane ac--field, it is to say that a linearly polarized ac--field
will have the Floquet TR symmetry. Furthermore, if Eq.\ref{linear} is held,
we can always choose $\tau _{0}=0$ and $\mathcal{Q}=\mathcal{IT}$. To show
this, let us shift the time frame $\tau =\tau ^{^{\prime }}+\tau _{0}/2$ and
plug in the Eq.\ref{sine_eq}. If so, we can get a new equation where $\phi
_{i}=n_{i}\pi $ and $\tau _{0}^{^{\prime }}=0$.
\end{document}